\documentclass[12pt]{article}
\begin{document}

\author{H. N. Agiza and A. A. Elsadany\\
Department of Mathematics, Faculty of Science\\
Mansoura University, Mansoura 35516, P. O. Box 64, Egypt\\
E-mail: agizah@mans.edu.eg (H.N. Agiza)}
\title{Nonlinear dynamics in the Cournot duopoly game with heterogeneous
players}
\date{}
\maketitle

\begin{abstract}
We analyze a nonlinear discrete-time Cournot duopoly game, where players
have heterogeneous expectations. Two types of players are considered:
boundedly rational and naive expectations. In this study we show that the
dynamics of the duopoly game with players whose beliefs are heterogeneous,
may become complicated. The model gives more complex chaotic and
unpredictable trajectories as a consequence of increasing the speed of
adjustment of boundedly rational player. The equilibrium points and local
stability of the duopoly game are investigated. As some parameters of the
model are varied, the stability of the Nash equilibrium point is lost and
the complex (periodic or chaotic) behavior occurs. Numerical simulations is
presented to show that players with heterogeneous beliefs make the duopoly
game behaves chaotically. Also we get the fractal dimension of the chaotic
attractor of our map which is equivalent to the dimension of Henon map.%
\newline
\newline
{\bf Keywords:} Discrete dynamical systems; Cournot duopoly games; Complex
dynamics; Heterogeneous expectations
\end{abstract}

\section{Introduction}

An oligopoly is a market system which is controlled by a few number of firms
producing homogeneous products. The dynamic of oligopoly game is more
complex because oligopolist must consider not only the behaviors of the
consumers, but also the reactions of the other competitors. Cournot, in 1838
[1] introduced the first formal theory of oligopoly, who treated the case
with naive expectations, so that in every step each player assumes the last
values taken by the competitors without estimation of their future reactions.

Recently several works have shown that the Cournot model may lead to complex
behaviors such as cyclic and chaotic, see for example {\rm [2, 3, 4, 5, 6,
7].} Among the first to do this was Puu [3, 4] who found a variety complex
dynamics arising in the Cournot duopoly case including the appearance of
attractors with fractal dimension. Others study of the dynamics of oligopoly
models with more firms and other modifications include Ahmed and Agiza [8],
Agiza [5] and Agiza et al. [9] such efforts have been extended by Bischi and
Kopel [10] in a duopoly game with adaptive expectations. The development of
complex oligopoly dynamics theory have been reviewed in [11].

Expectations play a key role in modelling economics phenomena. A producer
can choose his expectations rules of many available techniques to adjust his
production outputs. May be in the market of duopoly model each firm behaves
with different expectations strategies, so we are going to apply this kind
of expectations in our model which is common in reality{\it .}

In this paper we consider a duopoly model which is introduced in [7] but
each player form a different strategy in order to compute his expected
output. We take firm 1 represent a boundedly rational player while firm 2
has naive expectations. Each player adjusts his outputs towards the profit
maximizing amount as target and use his expectations rule. Recently examples
of oligopoly games with homogenous expectations are studied by Puu [4],
Kopel [6], Agiza [12], Agiza et al [13, 14].  It was shown that the dynamics
of Cournot oligopoly game may never settle to a steady state, and in the
long run they exhibit bounded dynamic which may be periodic or chaotic.
Economic model with heterogeneous players is introduced see [15,16]. Also
the dynamics of heterogeneous two-dimensional cobweb model have been studied
by Onozaki et al, see [17].

The main aim of this work is to investigate the dynamic behaviors of a
heterogenous model representing two firms using heterogeneous expectations
rules. This mechanism was applied in cobweb model [17] and gave us a guide
to apply it in our study.

The paper is organized as follows. In section 2 we describe the evolution of
dynamical systems of players with heterogeneous expectations rules. In
section 3, the dynamics of a duopoly game with boundedly rational player and
naive player is modelled by a two dimensional map .The existence and local
stability of the equilibrium points of the nonlinear map are analyzed.
Complex dynamics of behavior occur under some changes of control parameters
of the model which are shown by numerical experiments. Fractal dimension of
the strange attractor of the map is measured numerically.

\section{Heterogeneous expectations}

In oligopoly game players choose simple expectations such as naive
or complex as rational expectations. The players can use the same
strategy (homogenous expectations) or use different strategies
(heterogeneous expectations). Several economic models represent
the dynamics of heterogeneous firms, have been proposed in recent
years see for example [17, 18]. In this study we consider duopoly
game where each player has different strategy to maximize his
profit.

Firms can use rational expectations if they assume perfect
knowledge of underlying market and this may not be available in
real economic market. Also it is well known that in a duopoly
model with a heterogeneous firms their outputs depend upon
expectations of all competitors. Hence rational expectations can
only be achieved under unrealistic assumptions. For this reason
firms try to use another and my more realistic method which is
called bounded rationality. Firms usually do not have a complete
knowledge of the market, hence they try to use partial information
based on the local estimates of the marginal profit. At each time
period $t$ each firm increases (decreases) its production
$q_{i}\,$at\thinspace the period$\,(t+1) $ if the marginal profit
is positive (negative). If the players use this kind of
adjustments then they are boundedly rational players.

Let us consider a Cournot duopoly game where $q_{i}$ denotes the quantity
supplied by firm $i,i=1,2$. In addition let $P(q_{i}+q_{j})$, $i\neq j,$
denote a twice differentiable and nonincreasing inverse demand function and
let $C_{i}(q_{i})$ denote the twice differentiable increasing cost function.
Hence the profit of firm $i$ is given by
\begin{equation}
\Pi _{i}=P(q_{i}+q_{j})q_{i}-C_{i}(q_{i})
\end{equation}

At each time period every player must form an expectation of the rival's
output in the next time period in order to determine the corresponding
profit-maximizing quantities for period $t+1$. If we denote by $q_{i}(t)$
the output of firm $i$ at time period $t$, then its production $%
q_{i}(t+1),\,i=1,2\,\,$for the period $t+1\ $is decided by solving
\thinspace the two optimization problems.
\begin{equation}
\left\{
\begin{array}{c}
q_{1}(t+1)=\arg\max_{q_{1}}\Pi _{1}(q_{1}(t),q_{2}^{e}(t+1)) \\
q_{2}(t+1)=\arg\max_{q_{2}}\Pi _{2}(q_{1}^{e}(t+1),q_{2}(t)),
\end{array}
\right.  \
\end{equation}
where the function $\Pi _{i}(.,.)$ denotes the profit of the $i$
the firm and $q_{j}^{e}(t+1)$ represents the expectation of firm
$i$ about the production
decision of firm $j,\,(j=1,2,j\neq i)$. Cournot [1] assumed that $%
q_{i}^{e}(t+1)=\,q_{i}(t)\,\,,$ firm $i$ expects that the production of firm
$j$ will remain the same as in current period ( naive expectations). The
solution of the optimization problem of producer $i$ can be expressed as$%
\,q_{1}(t+1)=f\,(q_{2}(t))$,and $q_{2}(t+1)=g(q_{1}(t))$, so that the time
evolution of the duopoly system is obtained by the iteration of the
two-dimensional map $T:R^{2}\rightarrow R^{2}$ given by

\begin{equation}
T:\left\{
\begin{array}{c}
q_{1}^{\prime }=f(q_{2}) \\
q_{2}^{\prime }=g(q_{1})
\end{array}
\right.  \
\end{equation}

where$^{\prime }$ represents the one-period advancement operator.

The map (3) describes the duopoly game in the case of homogenous
expectations (naive). The fixed points of the map (3) are located at the
intersections of the two reaction functions $q_{1}=f(q_{2})$ and $%
q_{2}=f(q_{1})\,\,$and \thinspace are called Cournot-Nash equilibria of the
two-players game.

Firms try to use more complex expectations such as bounded
rationality, hence
they try to use local information based on the marginal profit $\frac{%
\partial \Pi _{i}}{\partial q_{i}}$ . At each time period $t$ each firm
increases (decreases) its production $q_{i}\,$at\thinspace the period$%
\,(t+1) $ if the marginal profit is positive (negative). If the players use
this kind of adjustments then they are boundedly rational players and the
dynamical equation of this game has the form
\begin{equation}
q_{i}(t+1)=q_{i}(t)+\alpha _{i}q_{i}(t)\frac{\partial \Pi
_{i}}{\partial q_{i}(t)},t=0,1,2,..  \
\end{equation}

where $\alpha _{i}$ is a positive parameter which represents the
speed of adjustment. The dynamics of the game (4) was studied by
Bischi and Naimzada [7 ].

In duopoly game with players use a different expectations for example the
first player is boundedly rational player and the other is naive. Hence the
duopoly game in this case are composed from the first equation of (4) and
the second equation of (3). Thus the discrete dynamical system in this case
is described by:

\begin{equation}
\left\{
\begin{array}{c}
q_{1}(t+1)=q_{1}(t)+\alpha _{1}q_{1}(t)\frac{\partial \Pi
_{1}}{\partial
q_{1}(t)} \\
q_{2}(t+1)=g(q_{1}(t))
\end{array}
\right.  \
\end{equation}

Therefore Eq. (5) describes the dynamics of a duopoly game with two players
using heterogenous expectations. In the next section we are going to apply
this technique to a duopoly model with linear demand and cost functions.

\section{The model}

Let $q_{i}(t),\,i=1,2$ represents the output of $i$th supplier
during period $t$, with a production cost function $C_{i}(q_{i}).$
The price prevailing in period $t$ is determined by the total
supply $Q(t)=q_{1}(t)+q_{2}(t)$ through a linear demand function
\begin{equation}
P=f(Q)=a-bQ  \
\end{equation}

where $a$ and $b$ positive constants of demand function. The cost function
is taken in the linear form
\begin{equation}
C_{i}(q_{i})=c_{i}q_{i},\,\,\,\,\,\,i=1,2  \
\end{equation}

where $c_{i}$ is the marginal cost of $i$th firm. With these
assumptions the single profit of $i$th firm is given by
\begin{equation}
\Pi _{i}=q_{i}(a-bQ)-c_{i}q_{i},i=1,2  \
\end{equation}

Then the marginal profit of $i$th firm at the point
$(q_{1},q_{2})\,$of the strategy space is given by
\begin{equation}
\frac{\partial \Pi _{i}}{\partial q_{i}}=a-c_{i}-2bq_{i}-bq_{j},i,j=1,2,j%
\neq i  \
\end{equation}

This optimization problem has unique solution in the form
\begin{equation}
q_{i}=r_{i}(q_{j})=\frac{1}{2b}(a-c_{i}-bq_{j})  \
\end{equation}

If the two firms are naive players, then the duopoly game is
describe from Eq.(3) by using Eq. (10) which has a linear form and
the Nash equilibrium is asymptotically stable [6]. In this study
we consider two players with different expectation, which the
first is boundedly rationality player and the other naive player.
The dynamic equation of the first player (boundedly rational
player) is obtained from inserting (9) in (3) which has the form:

\begin{equation}
q_{1}(t+1)=q_{1}(t)+\alpha q_{1}(t)(a-c_{1}-2bq_{1}(t)-bq_{2}(t))
\
\end{equation}

Using second equation of Eq. (3) the second player (naive) updates his
output according to the dynamic equation:
\begin{equation}
q_{2}(t+1)=\frac{1}{2b}(a-c_{2}-bq_{1}(t))  \
\end{equation}

Then the duopoly game with heterogeneous players is described by a
two-dimensional nonlinear map
\[
T(q_{1},q_{2})\rightarrow (q_{1}^{\prime },q_{2}^{\prime })
\]

which is defined from coupling the dynamic equations (11) and (12) as
follows:
\begin{equation}
T:\left\{
\begin{array}{l}
q_{1}^{\prime }=q_{1}+\alpha q_{1}(a-c_{1}-2bq_{1}-bq_{2}) \\
q_{2}^{\prime }=\frac{1}{2b}(a-c_{2}-bq_{1})
\end{array}
\right.   \
\end{equation}

The map (13) is a noninvertable map of the plane. The study of the dynamical
properties of the map (13) allows us to have information on the long-run
behavior of heterogenous players. Starting from given initial condition $%
(q_{1_{0},}q_{2_{0}})$, the iteration of equation (13) uniquely determines a
trajectory of the states of firms output.

$(q_{1}(t),q_{2}(t))=T^{t}(q_{1_{0},}q_{2_{0}}),\,t\ =0,1,2,...$

\subsection{Equilibrium points and local stability}

The fixed points of the map (13) are obtained as nonnegative solutions of
the algebraic system
\begin{eqnarray*}
q_{1}(a-c_{1}-2bq_{1}-bq_{2}) &=&0 \\
(a-c_{2}-2bq_{2}-bq_{1}) &=&0
\end{eqnarray*}

which is obtained by setting $q_{i}^{\prime }=q_{i},i=1,2$ in Eq. (13). We
can have at most two fixed points $E_{0}=(0,\frac{a-c_{2}}{2b})\,\,$and $%
E_{*}=(q_{1}^{*},q_{2}^{*}).$ The fixed point $E_{0}$ is called a boundary
equilibrium [7] and have economic meaning when $c_{2}<a$. The second
equilibrium $E_{*}$ is called Nash equilibrium where
\begin{equation}
q_{1}^{*}=\frac{a+c_{2}-2c_{1}}{3b}\,\;\rm{ and }\;\,q_{2}^{*}=\frac{a+c_{1}-2c_{2}%
}{3b}  \
\end{equation}
provided that
\begin{equation}
\left\{
\begin{array}{c}
2c_{1}-c_{2}<a \\
2c_{2}-c_{1}<a
\end{array}
\right.  \
\end{equation}

It is easy to verify that the equilibrium point $E_{*}$ is located at the
intersection of the two reaction curves which represent the locus of points
of vanishing marginal profit in Eq. (9). In the following, we assume that
Eq. (15) is satisfied, so the Nash equilibrium $E_{*}$ exists.

The study of the local stability of equilibrium solutions is based on the
localization, on the complex plane of the eigenvalues of the Jacobian matrix
of the two dimensional map (Eq. (13)).

The study of the local stability of equilibrium solutions is based on the
localization, on the complex plane of the eigenvalues of the Jacobian matrix
of the two dimensional map (Eq. (13)).

The Jacobian matrix of the map (13) at the state $(q_{1},q_{2})\,$has the
from:
\begin{equation}
J(q_{1},q_{2})=\left[
\begin{array}{cc}
1+\alpha (a-4bq_{1}-bq_{2}-c_{1}) & -\alpha bq_{1} \\
\frac{-1}{2} & 0
\end{array}
\right]  \
\end{equation}

The determinant of the matrix $J$ is
\[
Det=-\frac{1}{2}\alpha bq_{1}
\]
Hence the map (13) is dissipative dynamical system when $\left| \alpha
bq_{1}\right| <2.$

{\bf Lemma 1:\thinspace \thinspace \thinspace } The fixed point $E_{0}$ of
the map (Eq. (13)) is unstable .

{\bf Proof: \thinspace \thinspace \thinspace } In order to prove this
results, we estimate the eigenvalues of Jacobian matrix\thinspace $J$ at $%
E_{0}$. The Jacobian matrix has the form
\[
J(E_{0})=\left[
\begin{array}{cc}
1+\frac{\alpha }{2}(a-2c_{1}+c_{2}) & 0 \\
-\frac{1}{2} & 0
\end{array}
\right]
\]

The matrix $J(E_{0})\,\,$has two eigenvalues $\lambda _{1}=1+\frac{\alpha }{2%
}(a-2c_{1}+c_{2})$ and $\lambda _{2}=0.$ From condition (15), it follows
that $\left| \lambda _{1}\right| >1$. Then $E_{0}$ is unstable fixed point
(saddle point) for the map (13) and this completes the proof

\subsubsection{Local stability of Nash equilibrium}

We study the local stability of Nash equilibrium of two-dimensional map
(13). The Jacobian matrix (16) at $E_{*}$, which take the form
\begin{equation}
J(E_{*})=\left[
\begin{array}{ll}
1-2\alpha bq_{1}^{*} & -\alpha bq_{1}^{*} \\
-\frac{1}{2} & 0
\end{array}
\right]  \
\end{equation}

The characteristic equations of $J(E_{*})$ is $P(\lambda )=\lambda
^{2}-Tr\lambda +Det=0$ where $Tr$ is the trace and $Det$ is the determinant
of the Jacobian matrix defines in (17), $Tr=1-2\alpha bq_{1}^{*}$ and $Det=-%
\frac{1}{2}\alpha bq_{1}^{*},$

Since $(Tr)^{2}-4Det=(1-2\alpha bq_{1}^{*})^{2}+2\alpha bq_{1}^{*}$. It is
clear that $(Tr)^{2}-4Det>0$ ( has positive discriminant), then we deduce
that the eigenvalues of Nash equilibrium are real. The local stability of
Nash equilibrium is given by using Jury's conditions [21] which are:

\begin{enumerate}
\item  $\left| Det\right| \,<1$

\item  $1-Tr+Det>0,\,$and

\item  $1+Tr+Det>0$
\end{enumerate}

The first condition is $\left| \alpha bq_{1}^{*}\right| <2,$ which implies
that
\begin{equation}
\alpha <\frac{6}{(a-2c_{1}+c_{2})}  \
\end{equation}

The second condition $1-Tr+Det=\frac{3}{2}\alpha bq_{1}^{*}>0$, then the
second condition is satisfied. Then the third condition becomes:
\[
\frac{5}{2}\alpha bq_{1}^{*}-2<0
\]

This inequality is equivalent to
\begin{equation}
\alpha <\frac{12}{5(a-2c_{1}+c_{2})}  \
\end{equation}

Form (18) and (19) it follows that the Nash equilibrium is stable
if $\alpha <\frac{12}{5(a-2c_{1}+c_{2})}$ and hence the following
lemma is proved.

{\bf Lemma 2:\thinspace \thinspace \thinspace } The Nash equilibrium $%
E_{*}\, $of the map\ (Eq. (13)) is stable provided that $\alpha <\frac{12}{%
5(a-2c_{1}+c_{2})}.$

From the previous lemma, we obtain information of the effects of the model
parameters on the local stability of Nash equilibrium point\thinspace $E_{*}$%
. For example, an increase of the speed of adjustment of boundedly
rational player with the other parameters held fixed, has a
destabilizing effect. In fact, an increase of $\alpha $, starting
from a set of parameters which ensures the local stability of the
Nash equilibrium, can bring out the region of the stability of
Nash equilibrium point, crossing the flip bifurcation surface
\,\,\thinspace $\alpha =\frac{12}{5(a-2c_{1}+c_{2})}$. Similar
argument apply if the parameters $\alpha ,\;b,$ $c_{1}$ and
$c_{2}$ are fixed parameters and the parameter $a$, which
represents the maximum price of the good produced, is increased.
In this case the region of stability of $E_{*}$ becomes small, and
this implies that $E_{*}$ losses its stability. Complex behaviors
such as period doubling and chaotic attractors are generated where
the maximum Lyapunov exponents of the map (16) become positives.

\subsection{Numerical investigations}

The main purpose of this section is to show that the qualitative
behavior. of the solutions of the duopoly game (13 ) with
heterogeneous player generates a complex behavior that the case of
duopoly game with homogenous (naive) player.

To provide some numerical evidence for the chaotic behavior of system (13),
we present various numerical results here to show the chaoticity, including
its bifurcations diagrams, strange attractors, Laypunov exponents, Sensitive
dependence on initial conditions and fractal structure. In order to study
the local stability properties of the equilibrium points, it is convenient
to take the parameters values as follows: $a=10,b=.5,c_{1}=3$ and $c_{2}=5.$

The Figure 1 shows the bifurcation diagram with respect to the parameter $%
\alpha $ (speed of adjustment of boundedly rational player), while the other
parameters are fixed ( $a=10,b=.5,c_{1}=3,$ and $c_{2}=5).$ In fact a
bifurcation diagram of a two-dimensional map (13) shows attractor of the
model (13) as a multi-valued of two-dimensional map of one parameter. In
Figure 1 the bifurcation scenario is occurred, if $\alpha $ is small then
three exists a stable equilibrium point (Nash). As one can see the Nash
equilibrium point $(6,2)$ is locally stable for small values of $\alpha $.
As $\alpha $ increases, the Nash equilibrium becomes unstable, infinitely
many period doubling bifurcations of the quantity behavior becomes chaotic,
as $\alpha $ increased. It means for a large values of speed of adjustment
of bounded rational player $\alpha $, the system converge always to complex
dynamics. Also one can see that the period doubling bifurcation occur at $%
\alpha =\frac{4}{15}$. If the case of $\alpha >\frac{4}{15}$, one
observes
flip bifurcation occurs and complex dynamic behavior begin to appear for $%
\frac{4}{15}<\alpha <1.$

A bifurcation diagram with respect to the marginal cost of the first player $%
c_{1}$, while other parameters are fixed as follow $a=10,b=.5,$ $c_{2}=5$
and $\alpha =.335$, is shown in Figure 2.

We show the graph of a strange attractor for the parameter constellation $%
(a,b,c_{1},c_{2},\alpha )=(10,.5,3,5,.42)$ in Figure 3, which exhibits a
fractal structure similar to Henon attractor [22].

In order to analyze the parameter sets for which aperiodic
behavior occurs, one can compute the maximal Lyapunov exponent
depend on $\alpha $. For example, if the maximal Lyapunov exponent
is positive, one has evidence for chaos. Moreover, by comparing
the standard bifurcation diagram in $\alpha $, one obtains a
better understanding of the particular properties of the system.
In order to study the relations between the local stability of the
Nash equilibrium point and the speed of adjustment of boundedly
rational player $\alpha $, one can compute the maximal Lyapunov
exponents for adjustment factor in the environment of $1$. Figure
4 displays the related maximal Lyapunov exponents as a function of
$\alpha $. From Figure 4, one can easily determine the degree of
the local stability for different values of $\alpha \in
(\frac{4}{15},1)$. At value of $\alpha >\frac{4}{15}$ the maximal
Lyapunov exponents is positive. A positive value of maximal
Lyapunov exponents implies sensitive dependence on initial
condition for chaotic behavior. From the maximal Lyapunov
exponents diagram it is easy to determine the parameter-sets for
which the system converges to cycles, aperiodic, chaotic behavior.
Beyond that its even possible to differentiate between cycles of
very higher order and aperiodic behavior of the map (13) see
Figure 4.

\subsubsection{Sensitive dependence on initial conditions}

To demonstrate the sensitivity to initial conditions of system (13), we
compute two orbits with initial points $(q_{1_{0}},q_{2_{0}})$ and $%
(q_{1_{0}}+.0001,q_{2_{0}})$, respectively. The results are show in Fig. 5
and Fig. 6. At the beginning the time series are indistinguishable; but
after a number of iterations, the difference between them builds up rapidly.

Figure 5 and figure 6 show sensitive dependence on initial conditions, $%
q_{1} $-coordinates of the two orbits, for system (13), plotted against the
time with the parameter constellation $(a,b,c_{1},c_{2},\alpha
)=(10,.5,3,5,.4)$; the $q_{1}$-coordinates of initial conditions differ by $%
0.0001$, the other coordinate kept equal.

\subsubsection{Fractal dimension of the map (13)}

Strange attractors are typically characterized by fractal dimensions. We
examine the important characteristic of neighboring chaotic orbits to see
how rapidly they separate each other. The Lyapunov dimension see [23] is
defined as follows:
\[
d_{L}=j+\frac{\sum_{i=1}^{i=j}\lambda _{i}}{\left| \lambda _{j}\right| }
\]

with $\lambda _{1},\lambda _{2},....,\lambda _{n}$, where $j$ is the largest
integer such that $\sum_{i=1}^{i=j}\lambda _{i}\geq 0$ and $%
\sum_{i=1}^{i=j+1}\lambda _{i}<0.$

In our case of the two dimensional map has the Lyapunov dimension which is
given by
\[
d_{L}=1+\frac{\lambda _{1}}{\left| \lambda _{2}\right| },\,\,\,\,\,\,\,\,%
\lambda _{1}>0>\lambda _{2}
\]

By the definition of the Lyapunov dimension and with help of the computer
simulation one can show that the Lyapunov dimension of the strange attractor
of system (13). At the parameters values $(a,b,c_{1},c_{2},\alpha
)=(10,.5,3,5,.41)$ two Lyapunov exponents exists and are $\lambda _{1}=0.28$
and $\lambda _{2}=-1.06.$ Therefore the map (13) has a fractal dimension $%
d_{L}\approx 1+\frac{0.28}{1.06}\approx 1.26$, which is the same
fractal dimension of Henon map [24].

\section{Conclusions}

We have investigated the dynamics of a nonlinear, two-dimensional
duopoly game, which contains two-types of heterogeneous
players:{\it \ }boundedly rational player and naive player. This
game is described by a two-dimensional noninvertible map. The
stability of equilibria, bifurcation and chaotic behavior are
analyzed. The influence of the main parameters (such as the speed
of adjustment of boundedly rational player, the maximum price of
demand function and the marginal costs of players) on the local
stability is studied. We deduced that introducing heterogeneous
expectations for players in the duopoly game cause a market
structure to behave chaotically.

${\bf Acknowledgment:}${\it \ }We thank T. Onazaki for his comments.

\bigskip

\bigskip \bigskip

\bigskip

\bigskip \bigskip \bigskip

\bigskip \bigskip

\bigskip

\bigskip \bigskip

\smallskip \newpage

\begin{center}
{\Large Figure Captions}
\end{center}

\begin{quotation}
{\bf Fig.1. }Shows the bifurcation diagram with respect to the parameter $%
\alpha $ speed of adjustment of bounded rational player, with other fixed
parameters $a=10,b=.5,c_{1}=3,$ and $c_{2}=5.$\newline

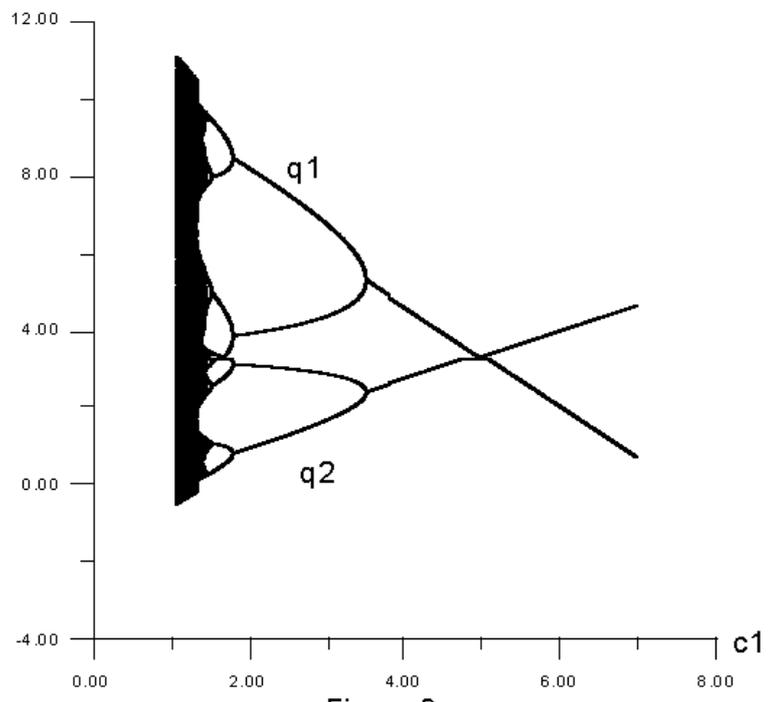
{\bf Fig.2. }A bifurcation diagram with respect to the marginal cost of the
first player $c_{1}$, with other parameters fixed at $a=10,b=.5,$ $c_{2}=5$
and $\alpha =.335$\newline

{\bf Fig.3. }We show the graph of a strange attractor for the parameter
constellation $(a,b,c_{1},c_{2},\alpha )=(10,.5,3,5,.41)$.\newline

{\bf Fig.4.} displays the related maximal Lyapunov exponents as a function
of $\alpha $.\newline

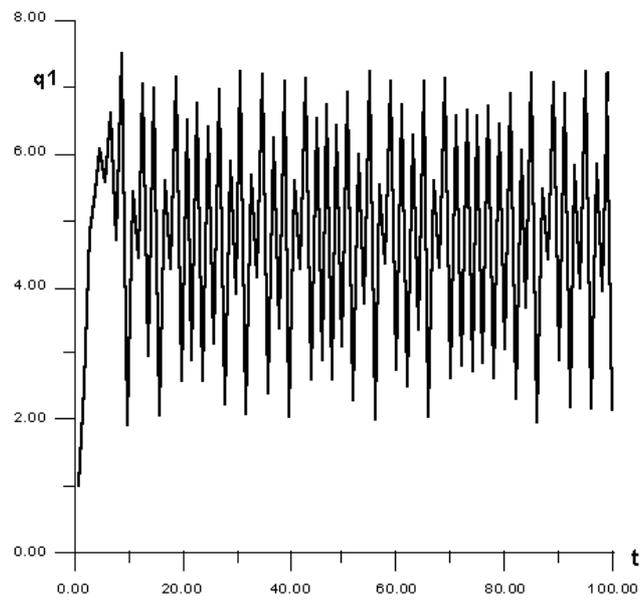
{\bf Fig.5. and Fig 6} show sensitive dependence on initial conditions, $%
q_{1}$-coordinates of the two orbits, for system (12), at the parameter
values $(a,b,c_{1},c_{2},\alpha )=(10,.5,3,5,.4)$;\newline
\end{quotation}

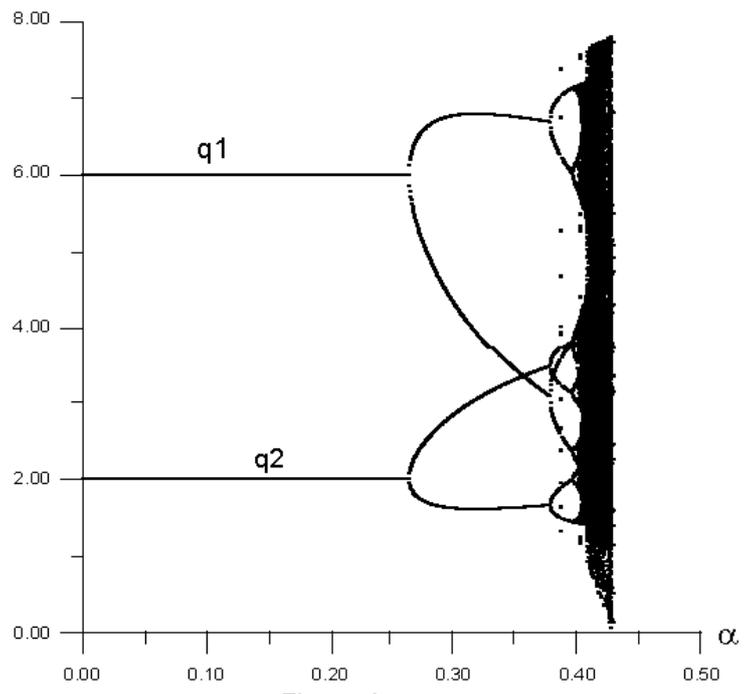
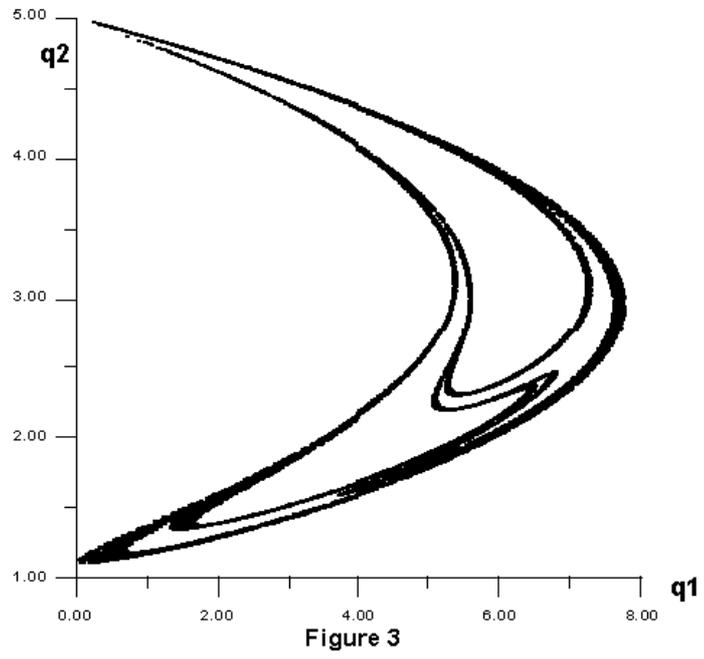
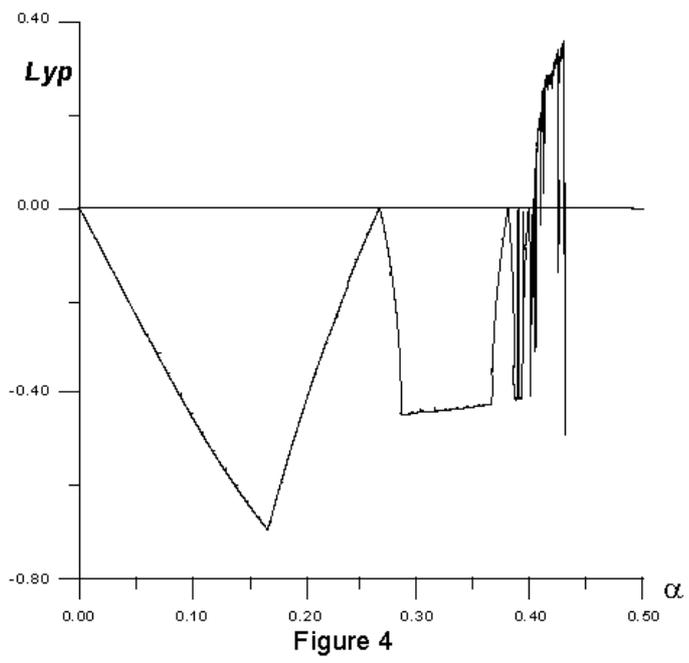
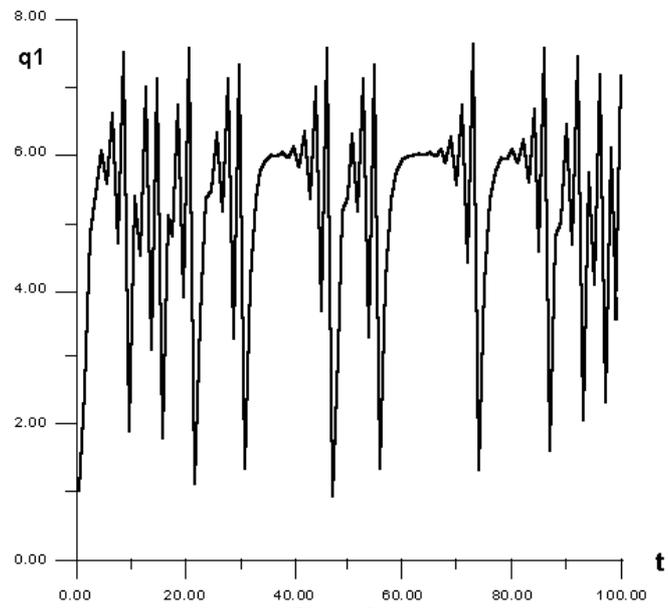
\end{document}